\documentclass[journal]{IEEEtran}

\usepackage[nocompress]{cite}
\usepackage{amsmath,amssymb,amsfonts}
\usepackage{algorithmic}
\usepackage{algorithm}
\usepackage{graphicx}
\usepackage{textcomp}
\usepackage{subcaption}
\usepackage{url}
\usepackage{multirow}
\usepackage{booktabs}
\usepackage[dvipsnames]{xcolor}
\begin{document}

\title{Study of Iterative Detection and Decoding for {Mixed Near- and Far-Field} XL-MIMO Systems}

\author{Nayyab~Haider~and~Rodrigo~C.~de~Lamare, Fellow, IEEE } 

\maketitle

\begin{abstract}
Extra-large multiple-input multiple-output (XL-MIMO) systems {serving users across near-field and far-field regions} experience spherical wavefront propagation that provides enhanced spatial resolution over traditional far-field systems. In this work, we propose a novel weighted rate (WR) ordering-based  successive interference cancellation (SIC) scheme that exploits the spatial degrees of freedom inherent in near-field propagation. We also develop an iterative detection and decoding (IDD) framework that integrates the proposed WR ordering with low-density parity-check (LDPC) coding and channel estimation. We then analyze the information rates and the impact of near-field channel characteristics. Numerical results show that the proposed WR-SIC and the IDD scheme outperform existing approaches. 
\end{abstract}

\begin{IEEEkeywords}
Near-field communications, XL-MIMO systems, interference cancellation, iterative detection and decoding.
\vspace{-1em}
\end{IEEEkeywords}

\section{Introduction}
\IEEEPARstart{T}{he} evolution toward sixth-generation (6G) wireless systems has driven significant interest in extra-large multiple-input multiple-output (XL-MIMO) systems \cite{jvece,snsxl,losxl}, where massive antenna arrays with apertures reaching several meters enable unprecedented spatial multiplexing capabilities beyond massive multiple-input multiple-output (MIMO) systems \cite{mmimo,wence}. In such systems, the conventional far-field assumption, which models electromagnetic propagation as plane waves, breaks down for users within the Rayleigh distance \cite{yuanwei23,cui2022near}. These users experience near-field propagation characterized by spherical wavefronts, creating spatial non-stationarity and distance-dependent phase variations across the antenna array. While this challenges traditional signal processing techniques, it offers enhanced spatial discrimination capabilities that can be exploited for interference mitigation \cite{zhang2023mixed, zhou2026mixed}.

The fundamental distinction between near-field and far-field propagation has key implications for system design. In near-field scenarios, the channel matrix exhibits higher effective rank due to the spherical wavefront characteristics, providing additional spatial degrees of freedom that can be leveraged for interference cancellation \cite{lu2024tutorial}. However, existing successive interference cancellation (SIC) techniques, including channel norm-based \cite{cho2010mimo, liu2016energy} and signal-to-interference-plus-noise ratio (SINR)-based approaches \cite{kim2006log, fodor2010impact}, do not explicitly account for near-field properties. {Related work 
has been studied for downlink multiple access schemes \cite{ding2023noma, ding2024resolution,11079717}
, but not for rate-weighted uplink MMSE-SIC.} Furthermore, receivers must address imperfect channel state information (CSI), which is particularly challenging in near-field scenarios due to the need for accurate distance and angle estimation \cite{cui2022channel, han2020channel}.

Near-field channel modeling has characterized spatial non-stationarity and visibility regions \cite{tang2024joint, yuan2022spatial}, and developed a hybrid-field channel estimation technique \cite{wei2021channel}. Iterative detection and decoding (IDD) schemes demonstrated significant performance gains when combined with SIC in various MIMO scenarios \cite{uchoa2015iterative, boutros2002iterative}. However, prior work lacks integrated frameworks that exploit near-field spatial characteristics through specialized ordering metrics and address practical interference mitigation specific to near-field propagation.

In this work, we propose a novel weighted rate (WR) ordering for SIC that weights each spatial stream's information rate by its squared singular value \cite{nfidd}.  
\textcolor{black}{Unlike channel-norm and capacity-based ordering, the proposed WR metric couples each stream's rate to its own singular value, which favors near-field users through their larger channel gains resulting from spherical wavefront propagation and targets the per-user reliability controlled by the decoding order.} We also develop an iterative detection and decoding (IDD) scheme that exploits the proposed WR-SIC receiver with LDPC codes and iterative processing. The IDD scheme employs soft information exchange between the WR-SIC detector and LDPC decoder through log-likelihood ratio (LLR) message passing, with extrinsic information feedback to update a priori probabilities iteratively. Simulations evaluate the WR-SIC and the IDD scheme against existing techniques.

\section{System Model}

We consider the uplink of a near-field XL-MIMO system where a base station (BS) equipped with $M$ receive antennas serves $K$ users, each equipped with $N_U$ transmit antennas. The BS employs a uniform linear array (ULA) with antenna spacing $d = \lambda/2$, where $\lambda = c/f_c$ is the wavelength corresponding to carrier frequency $f_c$ and $c$ is the speed of light. Each user transmits independent data streams from their antennas, resulting in $K \times N_U$ spatial streams that must be simultaneously detected at the BS. 

The electromagnetic field surrounding the antenna array can be divided into near-field and far-field regions based on the Rayleigh distance: \vspace{-0.75em}
\begin{equation}
R = \frac{2D^2}{\lambda}
\label{eq:rayleigh}
\end{equation}
where $D$ is the maximum array aperture. Users at distances $r < R$ operate in the near-field region with spherical wavefront propagation, while users at $r \geq R$ experience far-field plane wave characteristics.  
\vspace{-1.25em}

\subsection{Near-Field and Far-Field Channel Models}\vspace{-0.5em}

For users in the near-field region, the channel coefficient between the $m$-th transmit antenna of user $k$ and the $n$-th BS receive antenna follows the {non-uniform spherical-wave model of \cite{lu2024tutorial} with the Friis path-loss gain \cite{friis1946note}:} \vspace{-0.15em}
\begin{equation}
h_{n,m,k} = {\sqrt{\beta_{n,m,k}}\, g_{n,m,k}}\, e^{-j \frac{2\pi}{\lambda} d_{n,m,k}}
\label{eq:nf_channel}
\end{equation}
{where $g_{n,m,k} \sim \mathcal{CN}(0,1)$ is the small-scale fading coefficient, so that $\mathbb{E}[|h_{n,m,k}|^2] = \beta_{n,m,k}$ is consistent with the Friis power gain.} The path loss coefficient is: 

\begin{equation}
\beta_{n,m,k} = \frac{\lambda^2}{(4\pi)^2 d_{n,m,k}^2}
\label{eq:path_loss}
\end{equation}

The Euclidean distance is:
\begin{equation}
d_{n,m,k} = \sqrt{\sum_{i \in \{x,y,z\}} (i_{tx,k,m} - i_{rx,n})^2}
\label{eq:distance}
\end{equation}
where $i_{tx,k,m}$ and $i_{rx,n}$ denote antenna coordinates. 
For far-field users, the channel coefficients are expressed by the plane-wave approximation:
\begin{equation}
h_{n,m,k} = {\sqrt{\beta_k}\, g_{n,m,k}}\, e^{-j \frac{2\pi}{\lambda} \mathbf{k}_k \cdot \mathbf{r}_n}
\label{eq:ff_channel}
\end{equation}
where the path loss coefficient is
$\beta_k = \frac{\lambda^2}{(4\pi)^2 r_k^2}$
 with $r_k$ being the average distance from user $k$ to the BS, and the wave vector $\mathbf{k}_k \in \mathbb{R}^{3 \times 1}$ encoding the direction of arrival. \vspace{-0.75em}

\subsection{Signal Model and Spatial Degrees of Freedom}
We consider a hybrid network with $K_{NF}$ near-field and $K_{FF}$ far-field users, where $K_{NF} + K_{FF} = K$. The $M \times KN_U$ channel matrix is given by
\begin{equation}
\mathbf{H} = [\mathbf{H}_1, \mathbf{H}_2, \ldots, \mathbf{H}_K],
\label{eq:channel_matrix}
\end{equation}
where $\mathbf{H}_k \in \mathbb{C}^{M \times N_U}$ for user $k$. The received signal is: \vspace{-0.5em}
\begin{equation}
\mathbf{y} = \sum_{k=1}^{K} \mathbf{H}_k \mathbf{x}_k + \mathbf{n}
\label{eq:received_signal}
\end{equation} 
where $\mathbf{y} \in \mathbb{C}^{M \times 1}$ is the received vector, $\mathbf{x}_k \in \mathbb{C}^{N_U \times 1}$ contains transmitted symbols from user $k$ with total power $P_k$, and $\mathbf{n} \sim \mathcal{CN}(0, \sigma_n^2 \mathbf{I}_{M})$ is additive white Gaussian noise.

A key feature distinguishing near-field from far-field propagation is the effective channel rank. For the complete channel matrix, we perform singular value decomposition (SVD):
\begin{equation}
\mathbf{H} = \mathbf{U} \boldsymbol{\Sigma} \mathbf{V}^*
\label{eq:svd}
\end{equation}
where $\boldsymbol{\Sigma} = \text{diag}(\sigma_1, \sigma_2, \ldots, \sigma_{\min(M, KN_U)})$. In near-field scenarios, the spherical wavefront and distance-dependent phase variations result in
\begin{equation}
\mathbb{E}[\text{rank}(\mathbf{H}_{\text{near-field}})] > \mathbb{E}[\text{rank}(\mathbf{H}_{\text{far-field}})]
\label{eq:rank_comparison}
\end{equation}
{This rank increase is due to the distance-dependent quadratic phase term $\pi n^2 d^2 \cos^2\theta_{k,m}/(\lambda r_{k,m})$ in the near-field array response, which vanishes in the far field. Two users at the same angle but different distances have linearly independent near-field channel vectors, whereas their far-field channel vectors are collinear. This distance-domain separability raises the effective rank over the user distribution \cite{yuanwei23,lu2024tutorial}.} \vspace{-0.5em}

\section{Proposed Weighted Rate-SIC}

SIC ordering strategies based on channel norm \cite{cho2010mimo} or instantaneous SINR \cite{kim2006log} do not explicitly exploit the enhanced spatial structure of near-field channels. Since near-field users experience spherical wavefront propagation and distance variations across the array, they tend to have larger channel gains represented by larger singular values in their channel matrices.
For user $k$, we obtain the singular value decomposition (SVD):
\begin{equation}
\mathbf{H}_k = \mathbf{U}_k \boldsymbol{\Sigma}_k \mathbf{V}_k^*,
\label{eq:user_svd}
\end{equation}
where $\boldsymbol{\Sigma}_k = \text{diag}(\sigma_{k,1}, \sigma_{k,2}, \ldots, \sigma_{k,N_U})$. The weighted information rate for user $k$ is described by
\begin{equation}
{M_k} = \sum_{i=1}^{N_U} \sigma_{k,i}^2 \log_2 \left(1 + \frac{P_k \sigma_{k,i}^2}{N_U \sigma_n^2} \right)
\label{eq:weighted_capacity}
\end{equation}
The logarithmic term is the information rate of the $i$-th stream, while the weighting by $\sigma_{k,i}^2$ emphasizes streams with stronger channel gains. 
{Unlike channel-norm ordering $S_k^{CN}=\|\mathbf{H}_k\|_F^2=\sum_{i=1}^{N_U}\sigma_{k,i}^2$ \cite{cho2010mimo,liu2016energy}, which is rate-blind, and capacity-based ordering $S_k^{CAP}=\sum_{i=1}^{N_U}\log_2\!\big(1+P_k\sigma_{k,i}^2/(N_U\sigma_n^2)\big)$, which accounts for rate but weights all streams equally, the proposed metric weights each stream's rate by its own squared singular value. The function $f(\sigma^2)=\sigma^2\log_2(1+\rho\,\sigma^2)$ is convex and superlinear in $\sigma^2$, so ${M}_k$ is amplified for near-field users, whose spherical wavefront propagation yields larger and more evenly distributed singular values, even when their total channel power equals that of a far-field user, a case in which the channel norm cannot distinguish them. The metric also differs from asymptotic rate-scaling analysis such as the diversity-multiplexing tradeoff } \cite{zheng2003diversity} \textcolor{black}{ as it is designed specifically for stream ordering in SIC.}
\vspace{-1em}
\subsection{Ordering Algorithm and Transmission Scheme}
The ordering operates at two levels. At the user level, compute ${M_k}$ in \eqref{eq:weighted_capacity} for all users and sort in descending order:
\begin{equation}
{M_{k_1} \geq M_{k_2} \geq \cdots \geq M_{k_K}}
\label{eq:user_ordering}
\end{equation}
yielding user detection order $\pi_{\text{user}} = [k_1, k_2, \ldots, k_K]$.

At the stream level, for each user $k$, streams are ordered by descending singular values:
\begin{equation}
\sigma_{k,1} \geq \sigma_{k,2} \geq \cdots \geq \sigma_{k,N_U}
\label{eq:stream_ordering}
\end{equation}
The complete detection order processes all $N_U$ streams of user $k_1$ (ordered by decreasing singular value), then all streams of user $k_2$, creating:
\begin{equation}
\pi = [(k_1,1), (k_1,2), \ldots, (k_1,N_U), (k_2,1), \ldots, (k_K,N_U)]
\label{eq:combined_ordering}
\end{equation}
Each user $k$ transmits independent data streams from its $N_U$ antennas. The transmitted signal is given by
\begin{equation}
\mathbf{x}_k = [x_{k,1}, x_{k,2}, \ldots, x_{k,N_U}]^T
\label{eq:transmitted_signal}
\end{equation}
where each $x_{k,n_u}$ is drawn from a unit-energy constellation with power $P_k/N_U$ per antenna.
\vspace{-1em}
\subsection{Proposed WR-SIC Algorithm}

The WR-SIC processing algorithm initializes with $\mathbf{y}^{(1)} = \mathbf{y}$ and $\mathcal{R}^{(1)} = \{1, 2, \ldots, K \cdot N_U\}$, where $\mathbf{y}^{(1)}$ is the received vector for the first SIC stage and $\mathcal{R}^{(1)}$ represents the set of all $K \cdot N_U$ data streams to be detected. For each detection stage $i = 1, 2, \ldots, K \cdot N_U$, the algorithm proceeds as follows.

The stream to be detected at stage $i$ is selected according to the ordering from (\ref{eq:combined_ordering}):
\begin{equation}
(k_i^*, n_{u,i}^*) = \pi(i)
\label{eq:stream_selection}
\end{equation}
where $(k_i^*, n_{u,i}^*)$ identifies the user and antenna stream with the highest priority among the remaining undetected streams.

The minimum mean-square error (MMSE) receive filter or an alternative receive filter \cite{jidf} for this stream is then applied to the residual received signal $\mathbf{y}^{(i)}$ to obtain the soft estimate:
\begin{equation}
\bar{x}_{k_i^*,n_{u,i}^*} = \mathbf{w}_{k_i^*,n_{u,i}^*}^H \mathbf{y}^{(i)}
\label{eq:mmse_filtering}
\end{equation}
where $\mathbf{w}_{k_i^*,n_{u,i}^*}$ is the corresponding column of ${\mathbf{W}}_{\text{MMSE}}$ in 
\begin{equation}
\mathbf{W}_{\text{MMSE}} = \mathbf{H}^H(\mathbf{H}\mathbf{H}^H + \sigma_n^2\mathbf{I}_M)^{-1}
\label{eq:mmse_filter}
\end{equation}

{The soft estimate is then mapped to the nearest constellation point by a symbol-by-symbol minimum-distance decision:}
\begin{equation}
{\hat{x}_{k_i^*,n_{u,i}^*} = \mathcal{Q}\!\left(\bar{x}_{k_i^*,n_{u,i}^*}\right),}
\label{eq:ldpc_decode}
\end{equation}
{where $\mathcal{Q}(\cdot)$ denotes the hard-decision quantizer that maps the MMSE soft estimate to the nearest constellation symbol.} 

The contribution of the detected symbol is then subtracted from the residual received signal as follows: 
\begin{equation}
\mathbf{y}^{(i+1)} = \mathbf{y}^{(i)} - \mathbf{h}_{k_i^*,n_{u,i}^*} {\hat{x}_{k_i^*,n_{u,i}^*}}
\label{eq:interference_cancel}
\end{equation}
where $\mathbf{h}_{k_i^*,n_{u,i}^*}$ is the channel vector for the detected stream. 
The detected stream is then removed from the set of remaining streams:
\begin{equation}
\mathcal{R}^{(i+1)} = \mathcal{R}^{(i)} \setminus \{(k_i^*, n_{u,i}^*)\}
\label{eq:update_set}
\end{equation}

\textcolor{black}{WR-SIC processes all streams in the order specified by $\pi$. We remark that under ideal successive cancellation the MMSE-SIC receiver attains the multiple-access sum-capacity for any decoding order. Therefore, a capacity-based  ordering cannot increase capacity. The role of the proposed metric is different: the decoding order governs which streams face residual interference at each stage, and hence the per-user reliability and the error propagation through the successive stages. By detecting first the streams with strong singular-value structure, which is characteristic of near-field users, WR-SIC ensures reliable cancellation that benefits all remaining streams and exploits the spatial degrees of freedom of the near-field.}

\vspace{-1em}

\section{Analysis}

In this section, we analyze the proposed WR-SIC approach in terms of information rate and outage probability. \vspace{-1.25em}

\subsection{Information Rate Analysis}

The achievable rate for the $i$-th detected stream is given by
\begin{equation}
R_i = \log_2(1 + \text{SINR}_i^{\text{eff}})
\label{eq:rate_per_stream}
\end{equation}
where the effective SINR after SIC is:
\begin{equation}
\text{SINR}_i^{\text{eff}} = \frac{P_{k_i^*}/N_U \cdot |\mathbf{w}_{k_i^*,n_{u,i}^*}^H \mathbf{h}_{k_i^*,n_{u,i}^*}|^2}{\sum_{j \in \mathcal{R}^{(i+1)}} \frac{P_j}{N_U} \cdot |\mathbf{w}_{k_i^*,n_{u,i}^*}^H \mathbf{h}_j|^2 + \sigma_n^2 \|\mathbf{w}_{k_i^*,n_{u,i}^*}\|^2}
\label{eq:sinr_eff}
\end{equation}
The system sum-rate is given by
\begin{equation}
R_{\text{sum}} = \sum_{i=1}^{K \cdot N_U} R_i
\label{eq:sum_rate}
\end{equation}
For near-field scenarios, the enhanced spatial degrees of freedom and the proposed weighted rate ordering lead to:
\begin{equation}
\mathbb{E}[R_{\text{sum}}^{\text{near-field}}] \geq \mathbb{E}[R_{\text{sum}}^{\text{far-field}}],
\label{eq:sum_rate_comparison}
\end{equation}
 where the expectation is over user distributions and channel realizations. \textcolor{black}{This gain stems from prioritizing users with better channel conditions.} WR-SIC needs an SVD that costs $\mathcal{O}(K M N_U^2)$ operations and the MMSE filter costs $\mathcal{O}(M (KN_U)^2)$. {The schemes employ MMSE receive filters and LDPC decoders with similar cost. CN-SIC has the lowest ordering cost $\mathcal{O}(K M N_U)$, SINR-SIC the highest $\mathcal{O}(K^2 M N_U)$ as it recomputes the ordering at each stage, and WR-SIC lies between them at $\mathcal{O}(K M N_U^2)$ from its one-time per-user SVD. This ordering adds about $6.3\%$ on top of the MMSE cost that every scheme already incurs, so the $3$--$4$~dB gain of WR-SIC comes at a minor one-time extra cost.} 
\vspace{-1em}

\subsection{Outage Probability Analysis}

We derive closed-form outage probabilities for the general 
case of $K$ users under WR-SIC ordering in the region 
$\mathcal{R}_A \triangleq \{(R_{NF}, R_{FF}) : R_{NF} - 
R_{FF} > \delta_{th}\}$, where $\delta_{th} > 0$ is a 
stability threshold, and $\mathcal{R}_B \triangleq 
\{(R_{NF}, R_{FF}) : |R_{NF} - R_{FF}| \leq \delta_{th}\}$ 
denotes the region of comparable weighted rates.

Consider the WR-SIC decoding order $\pi = [\pi(1), \pi(2), 
\ldots, \pi(K)]$, where $\pi(i)$ denotes the user decoded 
at stage $i$, sorted in descending order of weighted rates 
as in \eqref{eq:user_ordering}. The user decoded at stage $i$ 
faces residual interference from all users at later stages 
$\pi(i+1), \ldots, \pi(K)$ that have not yet been 
cancelled. The effective SINR at stage $i$ is:
\begin{equation}
\text{SINR}_{\pi(i)}^{eff} = 
\frac{\frac{P}{N_U}|\mathbf{w}_{\pi(i)}^H 
\mathbf{h}_{\pi(i)}|^2}
{\displaystyle\sum_{j=i+1}^{K} \frac{P}{N_U}
|\mathbf{w}_{\pi(i)}^H \mathbf{h}_{\pi(j)}|^2 
+ \sigma_n^2\|\mathbf{w}_{\pi(i)}\|^2}
\label{eq:sinr_general}
\end{equation}
where the interference vanishes when $i = K$, so the last decoded user faces only thermal noise.   
\textcolor{black}{Under the channel of \eqref{eq:nf_channel}, we have for user $\pi(i)$: $\mathbf{h}_{\pi(i)} = \sqrt{\beta_{\pi(i)}}\,\boldsymbol{\alpha}_{\pi(i)} \odot \mathbf{g}_{\pi(i)}$, where $\beta_{\pi(i)}$ is the large-scale gain of \eqref{eq:path_loss} at the user distance, with the element-wise amplitude variation neglected in the analysis, $\boldsymbol{\alpha}_{\pi(i)}$ is the unit-modulus phase vector, and $\mathbf{g}_{\pi(i)} \sim \mathcal{CN}(\mathbf{0},\mathbf{I}_M)$ is the Rayleigh fading vector. The filter output is $\mathbf{w}_{\pi(i)}^H\mathbf{h}_{\pi(i)} \sim \mathcal{CN}\!\big(0,\beta_{\pi(i)}\|\mathbf{w}_{\pi(i)}\|^2\big)$, so the effective channel gain $|\mathbf{w}_{\pi(i)}^H\mathbf{h}_{\pi(i)}|^2$ is exponentially distributed with mean $\eta_{\pi(i)} = \beta_{\pi(i)}\|\mathbf{w}_{\pi(i)}\|^2$, which is larger for near-field users due to shorter distance and spherical wavefront propagation.}

The outage 
event for user $\pi(i)$ with target rate 
$\bar{R}_{\pi(i)}$ is:
\begin{equation}
\mathcal{O}_{\pi(i)} = \left\{\text{SINR}_{\pi(i)}^{eff} 
< \gamma_{\pi(i)}\right\}, \quad 
\gamma_{\pi(i)} = 2^{\bar{R}_{\pi(i)}} - 1
\label{eq:outage_event_general}
\end{equation}
Defining the outage threshold for user $\pi(i)$ as:
\begin{equation}
\phi_{\pi(i)}^{WR} \triangleq 
\frac{N_U \gamma_{\pi(i)}}{\eta_{\pi(i)}}, 
\quad i = 1, 2, \ldots, K
\label{eq:phi_general}
\end{equation}
the individual outage probability of user $\pi(i)$ is:
\begin{equation}
P_{out,\pi(i)}^{WR} = 1 - e^{-\phi_{\pi(i)}^{WR}/\rho}, 
\quad i = 1, 2, \ldots, K
\label{eq:user_outage_general}
\end{equation}

\textcolor{black}{where $\rho = P/\sigma_n^2$. Conditioned on the decoding order and treating residual interference as Gaussian \cite{boutros2002iterative}, the  system outage probability 
follows from the product of individual survival probabilities of the independent per-user gains:}
\begin{equation}
P_{sys}^{WR} = 1 - \prod_{i=1}^{K} 
\left(1 - P_{out,\pi(i)}^{WR}\right) 
= 1 - \exp\!\left(
-\frac{1}{\rho}\sum_{i=1}^{K}
\phi_{\pi(i)}^{WR}\right)
\label{eq:sys_outage_general}
\end{equation}
\textcolor{black}{The product form holds because each effective gain depends only on the corresponding user's own small-scale fading vector, which is independent across users; unconditional independence of the outage events is not assumed.} Applying the high-SNR approximation $1 - e^{-x} \approx x$ shows that the system outage decreases at rate $1/\rho$, {corresponding to diversity order one, since the system fails whenever any single user fails. The advantage of WR-SIC is not a higher diversity order but a smaller outage coefficient $\sum_{i=1}^{K}\phi_{\pi(i)}^{WR}$.} When CN-SIC  places a far-field user at an early stage, that user faces strong  residual interference from undecoded near-field users, yielding $\phi_{\pi(i)}^{CN} \gg \phi_{\pi(i)}^{WR}$ and therefore $P_{sys}^{CN} > P_{sys}^{WR}$,  analytically confirming WR-SIC superiority for arbitrary $K$. Eq. \eqref{eq:sys_outage_general} shows that the system outage is independent of decoding order under equal power allocation. However, the ordering affects individual user outage probabilities in \eqref{eq:user_outage_general}. WR-SIC 
assigns near-field users with large $\eta_{\pi(i)}$ to early stages with more residual interference, 
but their large mean channel gains keep 
$\phi_{\pi(i)}^{WR}$ small, ensuring reliable detection 
at every stage. In contrast, CN-SIC may assign a 
far-field user with small $\eta_{FF}$ to an early stage, 
yielding a large threshold and degraded individual outage performance.

\vspace{-0.5em}
\section{Proposed Iterative Detection and Decoding}

The proposed IDD scheme integrates the WR-SIC with LDPC coding through iterative soft information exchange \cite{spa,mfsic,mbdf,idd1bit,cdidd,msgamp,llraps,comp,iddocl,icce,nfidd}. The key components include: 1) soft WR-SIC-MMSE detector generating log-likelihood ratios (LLRs), 2) LDPC decoder producing refined symbol estimates using belief propagation and 3) extrinsic information feedback from decoder to detector updating a priori probabilities. 
For soft information processing, we model each stream after MMSE receive filtering as:
\begin{equation}
\bar{x}_{\pi(i)} = \mu_{\pi(i)} x_{\pi(i)} + \xi_{\pi(i)}
\label{eq:scalar_model}
\end{equation}
where $\bar{x}_{\pi(i)} = \mathbf{w}_{\pi(i)}^H \mathbf{y}^{(i)}$ is the filter output, $\mu_{\pi(i)} = \mathbf{w}_{\pi(i)}^H \mathbf{h}_{\pi(i)}$ is the effective channel gain, and $\xi_{\pi(i)} \sim \mathcal{CN}(0, \sigma^2_{\pi(i)})$ represents residual interference plus noise. The residual variance is:
\begin{equation}
\sigma^2_{\pi(i)} = \sigma_n^2 \|\mathbf{w}_{\pi(i)}\|^2 + \sum_{j \in \mathcal{R}^{(i+1)}} \text{Var}(x_j) |\mathbf{w}_{\pi(i)}^H \mathbf{h}_j|^2
\label{eq:residual_variance}
\end{equation}
With channel estimation, this becomes:
\begin{equation}
\hat{\sigma}^2_{\pi(i)} = \sigma_n^2 \|\hat{\mathbf{w}}_{\pi(i)}\|^2 + \sum_{j \in \mathcal{R}^{(i+1)}} \text{Var}(x_j) |\hat{\mathbf{w}}_{\pi(i)}^H \hat{\mathbf{h}}_j|^2 + \epsilon_{\text{est}},
\label{eq:residual_variance_est}
\end{equation}

\textcolor{black}{where $\epsilon_{\text{est}}$ is the additional residual variance arising from the channel estimation error. Writing the estimated channel vector of the stream detected at stage $i$ as $\hat{\mathbf{h}}_{\pi(i)} = \mathbf{h}_{\pi(i)} + \mathbf{e}_{\pi(i)}$, the error perturbs the effective gain by $\Delta\mu_{\pi(i)} = \hat{\mathbf{w}}_{\pi(i)}^H \mathbf{e}_{\pi(i)} \sim \mathcal{CN}\!\left(0,\sigma_e^2\|\hat{\mathbf{w}}_{\pi(i)}\|^2\right)$, which gives $\epsilon_{\text{est}} = \sigma_e^2\,\|\hat{\mathbf{w}}_{\pi(i)}\|^2\, P/N_U$, where $\sigma_e^2$ is the channel estimation error variance.}

\vspace{-0.1em}
\subsection{LLR Computation and LDPC Decoding}

For QPSK modulation with Gray mapping, the LLR for bit $b_l$ of symbol $x_k$ is:
\begin{equation}
L_D(b_l) = \log \frac{\sum_{x \in \mathcal{X}^{+1}_l} \exp\left(-\frac{|\bar{x}_k - \mu_k x|^2}{\sigma^2_k}\right) P(x)}{\sum_{x \in \mathcal{X}^{-1}_l} \exp\left(-\frac{|\bar{x}_k - \mu_k x|^2}{\sigma^2_k}\right) P(x)} - L_C(b_l),
\label{eq:llr_computation}
\end{equation}
where $\mathcal{X}^{+1}_l$ and $\mathcal{X}^{-1}_l$ are constellation subsets, $L_C(b_l)$ is the extrinsic LLR of the decoder and the a priori probability is:
\begin{equation}
P(x) = \prod_{m=1}^{M_c} \frac{1}{1 + \exp(-x^{b_m} L_C(b_m))}
\label{eq:apriori_prob}
\end{equation}

\textcolor{black}{The LDPC decoder employs standard log-domain belief propagation with variable-node and check-node updates, and, as in \cite{uchoa2015iterative}, the extrinsic information fed back to the detector is
\begin{equation}
L_C^{(t)}(b_l) = L_{APP}^{(t)}(b_l) - L_D^{(t)}(b_l) = \sum_{c \in \mathcal{N}(v)} L_{c \to v}^{(t)}(b_l)
\label{eq:extrinsic_info}
\end{equation}}

\vspace{-1.25em}
\subsection{Iterative Algorithm with Channel Estimation}
The complete IDD algorithm initializes with $L_C^{(0)}(b_l) = 0$ (uniform priors) and performs initial SIC. The algorithm then iterates for $t = 1, 2, \ldots, T_{\text{IDD}}$ iterations, where $T_{\text{IDD}}$ is the maximum number of IDD iterations:

\begin{enumerate}
\item Update a priori probabilities using (\ref{eq:apriori_prob}) with $L_C^{(t-1)}$

\item Recompute residual signal:
\begin{equation}
\mathbf{y}^{(i,t)} = \mathbf{y} - \sum_{j < i} \mathbf{h}_{\pi(j)} \tilde{x}_{\pi(j)}^{(t-1)}
\label{eq:residual_update}
\end{equation}
\item Compute detector LLRs using (\ref{eq:llr_computation}) for all streams

\item LDPC decoder: perform $T_{\text{LDPC}}$ iterations, compute extrinsic LLRs, and obtain symbol estimates:
\begin{equation}
\tilde{x}_{\pi(i)}^{(t)} = \arg\max_{x \in \mathcal{S}} P^{(t)}(x)
\label{eq:refined_estimate}
\end{equation}
\vspace{-0.75em}
\item Check convergence: if all parity checks satisfied and $\|L_C^{(t)} - L_C^{(t-1)}\|_2 < {\epsilon_{th}}$, terminate, where ${\epsilon_{th}} = 10^{-3}$ is the convergence threshold.
\end{enumerate}
When operating with estimated channels, the IDD algorithm uses $\hat{\mathbf{H}}$ in place of $\mathbf{H}$. The estimation errors introduce additional uncertainty in the LLRs:
\begin{equation}
L_D^{\text{est}}(b_l) = L_D^{\text{perfect}}(b_l) \cdot (1 - {\kappa}) + \mathcal{N}(0, {\kappa} \sigma_L^2)
\label{eq:llr_degradation}
\end{equation}
where $\textcolor{black}{\kappa \propto \sigma_e^2}$. Despite this degradation, IDD is beneficial as the extrinsic information from LDPC decoding helps mitigate estimation errors through iterative refinement.

\vspace{-0.5em}
\section{Simulation Results}

\begin{figure}[h!]
\centering
\includegraphics[width=0.425\textwidth]{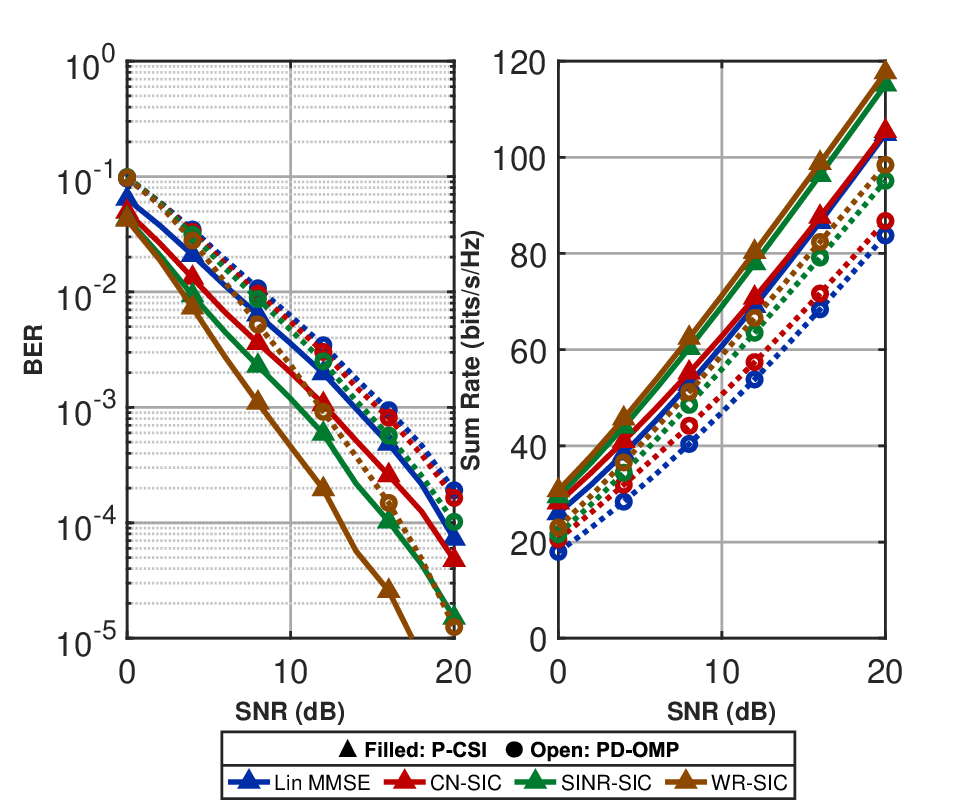}
 \vspace{-0.5em}
\caption{\small Average BER and sum rate with perfect and PD-OMP channel estimation. {Linear MMSE corresponds to the receiver of \cite{bacci2023spherical, lu2022multiuser}.}}
\label{fig: BER SR perfect and PD-OMP}
\end{figure}

{We evaluate the WR-SIC via simulations whose parameters are summarized in Table~\ref{tab:sim_params}.} {The near-field channel is generated according to the model of \cite{lu2024tutorial}.} The LDPC codes are constructed with the PEG algorithm \cite{peg,memd}, and channel estimation uses the polar-domain orthogonal matching pursuit (PD-OMP) algorithm~\cite{cui2022channel}. Two CSI conditions are evaluated: perfect CSI (P-CSI) and PD-OMP estimated CSI. Fig.~\ref{fig: BER SR perfect and PD-OMP} shows the  average BER (left) and sum-rate (right) under both CSI conditions. Under P-CSI, WR-SIC achieves a 3--4~dB SNR 
gain over CN-SIC and SINR-SIC at BER $= 10^{-3}$, while linear MMSE{, which corresponds to the near-field MMSE receiver of \cite{bacci2023spherical, lu2022multiuser}} performs the worst. WR-SIC also achieves the highest sum-rate across all SNR values. Under PD-OMP estimation, all schemes degrade, and WR-SIC 
consistently maintains its advantage, confirming robustness to imperfect CSI.

\begin{table}[t]
\centering
\caption{{Simulation Parameters}} \vspace{-0.5em}
\label{tab:sim_params}
{\begin{tabular}{ll}
\hline
\text{Parameter} & \text{Value} \\
\hline
Carrier freq. $f_c$ and wavelength $\lambda = c/f_c$            & 7.5 GHz and 0.04m  \\
BS antennas $M$ and array type                      & 64 and ULA\\
Antenna spacing $d$                  & $\lambda/2 = 0.02$ m \\
Array aperture $D = (M-1)d$        & 1.26 m \\
Rayleigh distance $R = 2D^2/\lambda$ & 79.38 m \\
Users $K$                            & 16 ($K_{NF}=8$, $K_{FF}=8$) \\
Antennas per user $N_U$              & 2 \\
Near-field user range $[1,\ 0.9R)$   & $[1,\ 71.44)$ m \\
Far-field user range $[1.1R,\ 2R]$   & $[87.32,\ 158.76]$ m \\
User azimuth $\theta$                & uniform in $\sin\theta\in[-1,1]$ \\
User placement                       & 2D azimuth--distance plane \\
Modulation                           & QPSK \\
SNR definition                       & $\rho = P/\sigma_n^2$ (Eq.~\eqref{eq:user_outage_general}) \\
Channel estimation                   & PD-OMP, pilot length $= 32$ \\
LDPC code rate and block length                    & 1/2  and  512 bits \\
LDPC (inner) and IDD (outer) iterations              & 20 and 3 \\
\hline \vspace{-1.25em}
\end{tabular}
}
\end{table}

\begin{figure}[h!]
\centering
\includegraphics[width=0.425\textwidth]{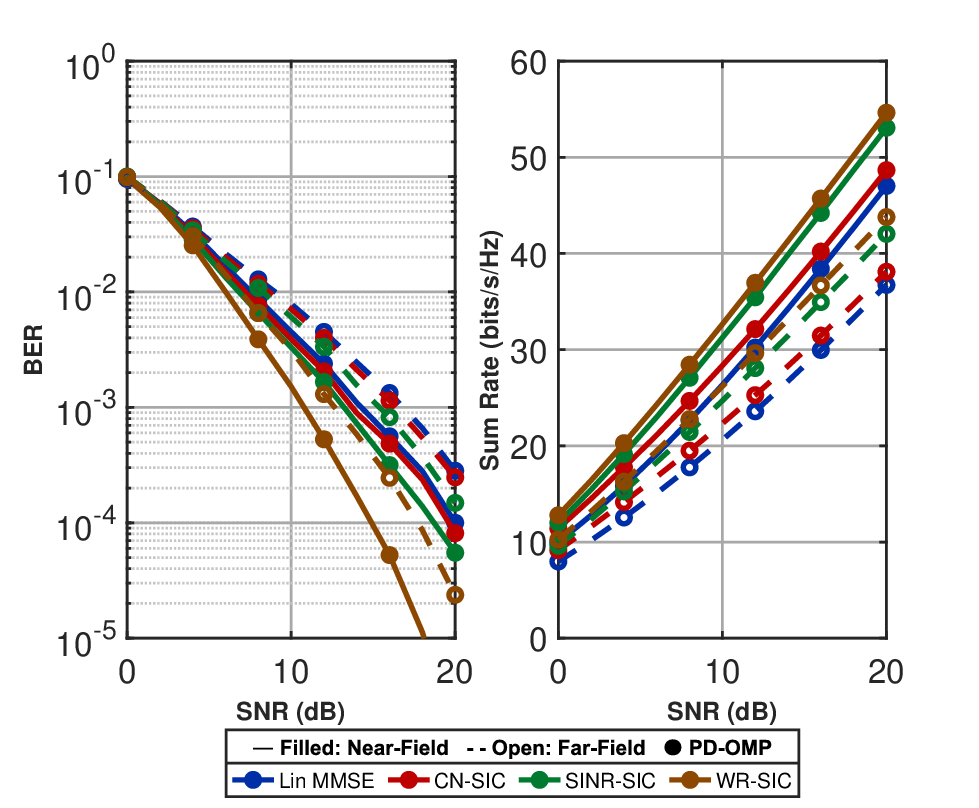}
 \vspace{-0.5em}
\caption{\small Comparison of near and 
far-field users with PD-OMP.} 
\label{fig: Near vs far-field PD-OMP}
\end{figure}

Fig.~\ref{fig: Near vs far-field PD-OMP} compares 
near-field and far-field users under PD-OMP estimation.  \textcolor{black}{Near-field users outperform far-field users for all schemes, while the gap widens at higher SNR.}
The sum-rate gap is most pronounced for WR-SIC, validating the rank enhancement 
in~(\ref{eq:rank_comparison}) and confirming that WR-SIC 
best exploits the near-field channel structure.

\begin{figure}[h!]
\centering
\includegraphics[width=0.425\textwidth]{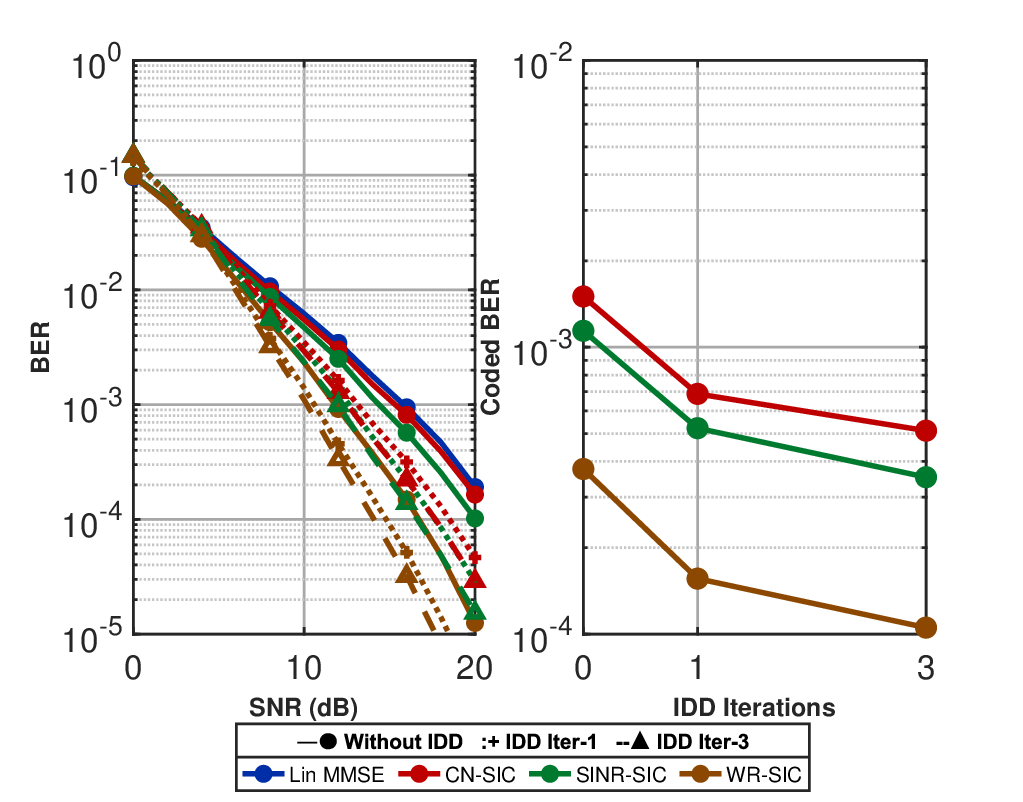}
 \vspace{-0.5em}
\caption{\small Coded BER performance of proposed and 
existing IDD schemes with PD-OMP channel estimation.}
\label{fig: with IDD}
\end{figure}

Fig.~\ref{fig: with IDD} presents the coded BER with IDD 
under PD-OMP estimation. WR-SIC achieves the largest 
absolute gain from iterative processing and converges 
faster to a lower BER floor than CN-SIC and SINR-SIC. 
The right subplot confirms that most available gain is 
captured within the first few iterations, demonstrating 
practical efficiency. The extra 2~dB gain from IDD further underscores the advantage of WR-SIC ordering in exploiting near-field spatial degrees of freedom. \vspace{-0.85em}

 \vspace{-0.05em}

 \vspace{-0.05em}

 \vspace{-0.05em}

\section{Conclusion}
We presented iterative interference cancellation techniques {for mixed near- and far-field XL-MIMO systems} with channel estimation. The proposed WR-SIC algorithm exploits the enhanced spatial degrees of freedom in near-field by weighting achievable rates with squared singular values. An IDD scheme integrating the proposed SIC with LDPC coding through soft information exchange was also developed. Numerical results demonstrate that WR-SIC achieves 3–4 dB SNR gain over conventional ordering schemes at BER = $10^{-3}$ under perfect CSI, with an extra 2 dB gain through IDD iterations. \vspace{-0.5em}

\bibliographystyle{IEEEtran} 
\bibliography{references}
\end{document}